\documentclass{article}
\usepackage{spconf,amsmath,amssymb,graphicx,hyperref,booktabs,dsfont, multirow,orcidlink}

\usepackage{comment}
\usepackage{multirow}
\usepackage{tikz}
\usetikzlibrary{shapes.geometric, arrows.meta, positioning, calc}

\newcommand{\iou}{\mathrm{IoU}}
\newcommand{\costmiss}{C_{\text{fr}}^{\classtar{}}}
\newcommand{\costfanon}{C_{\text{fa}}^{\classnon{}}}

\newcommand{\symbolhp}{{H}}

\newcommand{\classtar}{\mathsf{Y}}
\newcommand{\classnon}{\mathsf{N}}

\newcommand{\evidence}{\boldsymbol{x}}
\newcommand{\score}{s}

\newcommand{\acty}{\mathtt{ACCEPT}}
\newcommand{\actn}{\mathtt{REJECT}}

\title{SPOOF LOCALIZATION AS BOUNDARY REGRESSION}
\name{\begin{tabular}[t]{@{}c@{}}
      Yixuan Xiao$^{1}$\sthanks{Corresponding author: yixuan.xiao@ims.uni-stuttgart.de}\orcidlink{0009-0002-6272-8884}, Xin Wang$^{2}$\orcidlink{0000-0001-8246-0606}\sthanks{Thanks JST PRESTO Grant (JPMJPR23P9) for funding.},  Yassine El Kheir$^{3}$\orcidlink{0009-0003-3611-3154}, Feidi Kallel$^{3}$\orcidlink{0009-0005-7524-2922}, Arnab Das$^{3}$\orcidlink{0000-0001-9715-1435}, \\
      Tim Polzehl$^{3}$\orcidlink{0000-0001-9592-0296}, Sebastian M\"oller$^{4}$\orcidlink{0000-0003-3057-0760}, Ngoc Thang Vu$^{1}\orcidlink{0000-0001-7893-9147}$
      \end{tabular}}
\address{$^{1}$University of Stuttgart, Germany \quad
         $^{2}$National Institute of Informatics, Japan \\
         $^{3}$German Research Center for Artificial Intelligence (DFKI), Germany \\
         $^{4}$Technical University of Berlin, Germany}

\begin{document}
\ninept
\maketitle
\begin{abstract}
Partially spoofed audio embeds spoofed segments in a bonafide utterance, 
and spoof localization is the task of locating them. 
Most countermeasures (CMs) operate at the frame level. 
Typically, they include a front-end representing the audio as frames and a back-end classifying each frame as bonafide or spoof. 
Such frame-level detection ties the reference, the predictions and the evaluation to a pre-defined temporal resolution: a frame containing a transition receives a single label, a boundary cannot lie inside a frame, and the evaluation also depends on the resolution. 
To tackle these issues, we treat spoof localization as regression. 
We propose a single-stage boundary regressor that outputs $(\text{start-time}, \text{end-time}, \text{score})$ tuples with continuous-valued time stamps, trained on the exact reference intervals. 
Results show that the proposed CM outperforms previous frame-level and boundary-based CMs, and generalizes better.

\end{abstract}
\section{Introduction}
\label{sec:intro}
In the Partial Spoof (PS) scenario, an attacker does not synthesize a whole utterance 
but replaces a few words of a bonafide utterance with spoofed speech, so that the meaning changes while most of the audio stays bonafide~\cite{zhang2025partialedit, zhang2022partialspoof, luong2025llamapartialspoof}.
For such audio, an utterance-level decision is not always sufficient. In applications such as forensic analysis one might need to know which part of the utterance was edited. The task of locating the spoofed regions inside partially spoofed audio is known as spoof localization.

\begin{figure}[t]
\centering
\includegraphics[width=\columnwidth]{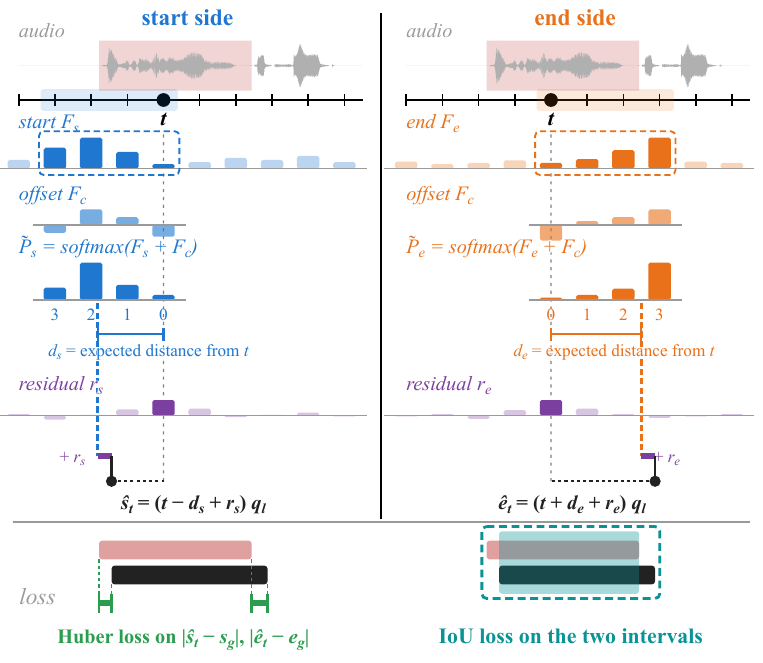}
\caption{Boundary decoding at one instant $t$ of one level. Start and end scores are predicted for all instants of that level; instant $t$ adds its own offset to the $B$ nearest scores on each side, converts them to a probability, and takes the expected distance plus a small residual as the boundary. The interval is trained with an IoU loss and each boundary with a Huber loss on its gap to the reference.}
\label{fig:method}
\end{figure}

Most countermeasures (CMs) for spoof localization treat the task as classification. A typical CM utilizes a self-supervised learning (SSL) model as the front-end~\cite{babu2022xlsr}, represents the audio as a sequence of feature vectors, called \emph{frames}, and uses a back-end that classifies each frame as bonafide or spoof. Note that the number of frames is set by the stride of the SSL convolutional layers; in wav2vec2-style front-ends~\cite{wu2024cfprf,rimon2026split, zhong2024bam, mao2026sal}, consecutive frames are 20~ms apart, and we call this distance the frame resolution.

However, the start and end times of a spoofed word are not necessarily multiples of 20~ms, whereas a frame-level CM can only see and predict on the 20~ms frame. This mismatch leads to three specific issues.
\textbf{First}, the reference labels must be quantized to the frame resolution, which introduces an unavoidable quantization error into training. 
More specifically, when partially spoofed audio is created, the spoofed interval $[t_s, t_e]$ is known exactly and is independent of the CM's temporal resolution. But to assign one label to each frame, a frame that is entirely spoofed and a frame of which only a small part is spoofed are assigned the same label ``spoof'', even though they have different levels of ``spoofness''.
\textbf{Second}, the precision of the model output is limited by the resolution, since a frame-level CM cannot predict a boundary inside a frame.
\textbf{Third}, the evaluation is also tied to the resolution: point-based EER changes markedly with the resolution at which frames are scored, so results are often reported at various resolutions, which makes comparison less straightforward. Although the evaluation issue can be alleviated by using the range-based EER~\cite{zhang2023rangeeer}, its theoretical foundation is not yet fully justified.
\footnote{See Appendix~\ref{app:rangeeer}.}
More importantly, it does not 
solve the first two issues. 

To tackle the above problems, we treat spoof localization as regression via relative boundary modeling. Inspired by TriDet~\cite{shi2023tridet}, we proposed a CM that takes an utterance as input and outputs a list of tuples $(s, e, c)$ with $s, e, c \in \mathbb{R}$, where $s$ and $e$ define a time range of a hypothesized spoof interval and $c$ is its spoof score.

Our \textbf{contributions}\footnote{\url{https://github.com/XIAOYixuan/boundary_detection/}} are: (1) a single-stage CM that regresses exact spoof boundaries, with no quantized label in training, prediction or evaluation. (2) On PartialSpoof the CM has the lowest point-based and range-based EER, and it generalizes better to LlamaPartialSpoof compared to other state-of-the-art methods. (3) The CM reveals a time misalignment problem in the released PartialEdit reference, and we released a corrected version using cross-correlation.

\section{Methods}
\label{sec: methods}

The proposed CM keeps the SSL front-end of a frame-level CM: a 300M XLS-R front-end represents the utterance as a sequence of 1024-dimensional frames. Its transformer blocks are fine-tuned while the convolutional feature extractor is frozen. 
What is predicted from these features changes in three aspects, as shown in Fig.~\ref{fig:arch}: (1) the frame sequence is expanded into a feature pyramid so that every position sees a different span of the input (\ref{ssec:pyramid}); (2) instead of one label per frame,  it regresses the exact start time and end time of the spoofed interval around each position at each level and predicts a spoof score of the interval (\ref{ssec:tasks}); (3) an utterance-level head acts as a gate to decide whether any spoofed interval should be output at all for an utterance (\ref{ssec:tasks}). 

We refer to one position of the resulting frame sequence at pyramid level $l$ as an \emph{instant} $t\in \{0, 1, \ldots, T_l-1\}$, where $T_l$ is the number of instants at that level. 
For every instant $t$ at every pyramid level $l$, the CM produces one \emph{proposal} $(\hat{s}_t, \hat{e}_t, \hat{c}_t)$. $\hat{c}$ is the spoof score of the interval defined by starting time $\hat{s}_t$ and end time $\hat{e}_t$:

\begin{equation}
\begin{split}
    \hat{s}_t &= \Delta \big[ (t - d_s + r_s)\, q_l \big] + \delta, \\
    \hat{e}_t &= \Delta \big[ (t + d_e + r_e)\, q_l \big] + \delta,
\end{split}
\label{eq: endpoint}
\end{equation}
where $\Delta$ is the frame resolution and 
$q_l$ is the stride at level $l$ (Sec.~\ref{ssec:pyramid}).
Although the values of $\hat{s}_t$ and $\hat{e}_t$ are largely
decided by $\Delta\,t\,q_l$, the additional shifting terms $d_s, d_e\in\mathbb{R}_{+}$ and $r_s, r_e\in \mathbb{R}$ help to predict the start time and end time more precisely. 
Among them, $d_s, d_e$ are the distances from time $\Delta\,t\,q_l$ to the actual start time or end time via probabilistic modeling (Sec.~\ref{ssec:tasks}); $r_s, r_e \in (-0.5, 0.5) $ are bounded residuals used to further refine each boundary by up to half a stride (Sec.~\ref{ssec:tasks}). Finally, 
$\delta$ is a constant, an implementation detail similar to the left padding in STFT.

\subsection{Multi-granular Modeling}
\label{ssec:pyramid}

\begin{figure}[t]
\centering
\includegraphics[width=\columnwidth]{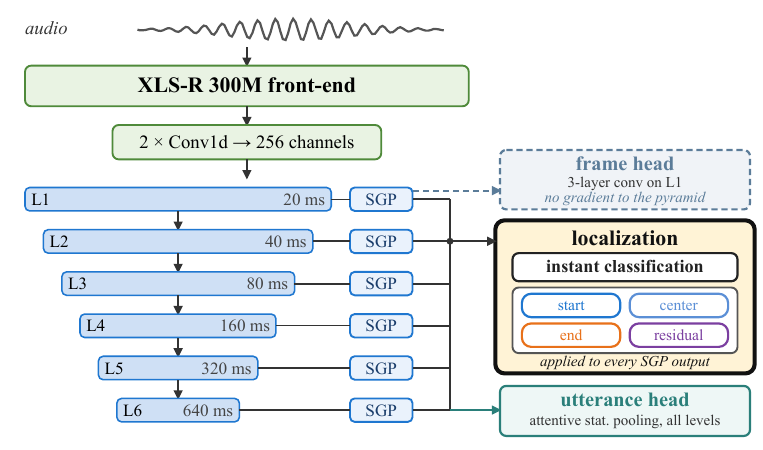}
\caption{Architecture overview.}
\label{fig:arch}
\end{figure}

This part follows the conventional frame-level CMs that build on XLS-R features:  
two 1-D convolutional modules with kernel size 3 are used to project the front-end output to $C = 256$ channels, 
and the sequence is then converted into a feature pyramid as in the multi-resolution CM~\cite{zhang2022partialspoof} which can capture fine-to-coarse features. As shown in Fig.~\ref{fig:arch}, the pyramid has six levels at 20 to 640~ms, each built from the previous one by max-pooling with stride 2, so that level $l$ has stride $q_l=2^{l-1}$ frames.

Each level is then processed by one SGP block proposed in~\cite{shi2023tridet}. The core function is defined as follows:

\begin{equation}
\begin{split}
F_\mathrm{SGP}(\mathbf{X}) = {}& \phi(\bar{\mathbf{x}}) \odot \mathrm{FC}(\mathbf{X}) \\
&+ \psi(\mathbf{X}) \odot \big(\mathrm{Conv}_w(\mathbf{X}) + \mathrm{Conv}_{kw}(\mathbf{X})\big) + \mathbf{X},
\end{split}
\label{eq:sgp}
\end{equation}
where $\mathbf{X}\in \mathbb{R}^{C\times T_l}$ is the feature sequence of the level with $T_l$ instants, and $\bar{\mathbf{x}} \in \mathbb{R}^{C}$ is its mean over the instants.
$\phi(\bar{\mathbf{x}})$ is computed once from the mean and applied to every instant, so each instant is weighted by the utterance-level context.  $\psi$, $\mathrm{Conv}_w$ and $\mathrm{Conv}_{kw}$ are depth-wise convolutions with two window sizes $w=1$ and $kw=3$, which model the local context of each instant. 
In each SGP block, the input is layer-normalized, and the output of Eq.~\eqref{eq:sgp} is followed by group normalization, a feed-forward network with a second residual connection, as in TriDet~\cite{shi2023tridet}. The resulting features are then fed into different heads to perform different tasks.

\subsection{Prediction tasks}
\label{ssec:tasks}
Let the reference of an utterance be a set of spoofed intervals $G = \{(s_g, e_g)\}$ in seconds, exact as released. 
A bonafide utterance has $|G| = 0$.
The three tasks of the proposed CM below are trained against these exact intervals; no reference is quantized. An auxiliary task, frame-level prediction, is added only for comparison with other frame-level CMs. It is the only task where quantized labels are used, and the gradient is \emph{stopped} at the prediction head, so the potential quantization error never reaches the proposed CM.

\noindent\textbf{Instant classification with center sampling.}
This task detects whether an instant is near the center of a reference spoofed interval. 
The head of this task is a three-layer convolutional module: the first two layers have kernel size 3, $C = 256$ channels and ReLU, and the last layer projects to one score.

Let $\mathbf{H}^l \in \mathbb{R}^{C \times T_l}$ denote the SGP block's output at level $l$, then for six levels there are six outputs $\mathbf{H}^1, \ldots, \mathbf{H}^6$. 
The head takes each $\mathbf{H}^l$ separately, and the same head is applied to all six levels. Its output has the same length $T_l$ as its input, which gives one spoof score $\hat{c}_t$ per instant. The same $\hat{c}_t$ is used as the spoof score of the interval predicted at $t$.
Each instant $t$ is labeled as positive (near the center) or negative (not near the center) by center sampling~\cite{tian2019fcos}, a method also used in~\cite{zhang2022actionformer, shi2023tridet}; i.e., center sampling assigns each positive instant to one reference spoofed interval $g$.
The head is trained with the binary focal loss $\mathcal{F}_t$~\cite{lin2017focal} as follows:

\begin{equation}
   \mathcal{L}_\mathrm{cls} =
   \frac{1}{|\mathcal{P}|} \sum_{t \in \mathcal{P}} \iou_t\, \mathcal{F}_t
   + \frac{1}{|\mathcal{N}|} \sum_{t \in \mathcal{N}} \mathcal{F}_t ,
\label{eq:cls}
\end{equation}
where $\mathcal{P}$ and $\mathcal{N}$ are the sets of positive and negative instants of a batch. $\iou_t$ is the IoU between the interval predicted at $t$ and the reference spoofed interval assigned to $t$. It weights the positives, so that $\hat{c}_t$ reflects how well the interval is localized~\cite{shi2023tridet}.

\noindent\textbf{Boundary regression.}
This task predicts the exact start and end of the spoofed interval for every positive instant $t$. 
According to Eq.~\eqref{eq: endpoint}, this requires the distances $d_s, d_e$ and the residuals $r_s, r_e$. As shown in Fig.~\ref{fig:method}, they are predicted by four heads: a start head, an end head, and an offset head together give the two distances, and a residual head give the two residuals. Each head has the same three-layer convolutional structure and receives the same SGP block's output as the instant classification head.

The input of the start and end heads is the same as the instant classification head: each head takes $\mathbf{H}^l$ as input, and the head is shared by all six levels. 
Distance prediction follows TriDet~\cite{shi2023tridet}. A start head and an end head output one score per instant, $F_s \in \mathbb{R}^{T_l}$ and $F_e \in \mathbb{R}^{T_l}$, which indicate how likely an instant is a start or an end.

The input of an offset head is also the same as the instant classification head. The offset head outputs $F_c \in \mathbb{R}^{T_l \times 2 \times (B+1)}$. For instant $t$, it adjusts the scores of $t$ and of its $B = 16$ neighbours on each side. 
Take the start time prediction for example: for $t$, the probability of it or its neighbours being the start $\tilde{P}_{s}(b)$ is $\mathrm{softmax}_{b}\big(F_s[t-b] + F_c[t, 0, b]\big)$, and the distance to start $d_s$ is the expectation $\sum_{b=0}^{B} b\, \tilde{P}_s(b)$.

A residual is slightly different from the other heads in this task: it has the same inputs, but only shares its first two layers. The residual corrects the error left by the distances, which differs between levels, so the last layer is separate for each level. About the error, more specifically, $d_s, d_e\ge 0$ force the predicted interval to contain $t$, while a spoofed interval shorter than the stride $q_l$ may start after $t$ or end before it. We therefore proposed a residual head to output $F_r \in \mathbb{R}^{T_l \times 2}$. The residuals $r=0.5\tanh(F_r[t])$ can push the start past $t$ to the right, or the end past $t$ to the left, by up to half a stride. 

With the distances and residuals, Eq.~\eqref{eq: endpoint} gives
the predicted interval $(\hat{s}_t, \hat{e}_t)$. Then we can train the heads by comparing the interval predicted at $t$ and the reference spoofed interval $g$ assigned to $t$.
As shown in Fig.~\ref{fig:method}, Huber loss $\mathcal{L}_\mathrm{hub}$ and IoU loss $\mathcal{L}_\mathrm{IoU}$ are adopted:

\begin{align}
   \mathcal{L}_\mathrm{hub} &= \frac{1}{|\mathcal{P}|} \sum_{t \in \mathcal{P}}
   \Big[ h\big(\tfrac{|\hat{s}_t - s_g|}{\tau_b}\big) + h\big(\tfrac{|\hat{e}_t - e_g|}{\tau_b}\big) \Big], \label{eq:hub} \\
   \mathcal{L}_\mathrm{IoU} &= \frac{1}{|\mathcal{P}|} \sum_{t \in \mathcal{P}} (1 - \iou_t), \label{eq:iou}
\end{align}
where $h(u) = u^2/2$ for $u < 1$ and $u - \tfrac{1}{2}$ otherwise
is the Huber function~\cite{meyer2021alternative}, $\tau_b = 20$~ms is the frame resolution, and $\iou_t$ is the IoU between $(\hat{s}_t, \hat{e}_t)$ and $g$. 
The two losses complement each other:  $\mathcal{L}_\mathrm{IoU}$ measures how well $(\hat{s}_t, \hat{e}_t)$ overlaps $g$, but an error of a few milliseconds barely changes it on a long interval. To make the prediction more precise, $\mathcal{L}_\mathrm{hub}$ penalizes the error at each boundary in absolute time, regardless of the interval length.

\noindent\textbf{Utterance-level Prediction.} 
This task predicts whether an utterance contains any spoofed intervals. So far, $\mathcal{L}_\mathrm{hub}$, $\mathcal{L}_\mathrm{IoU}$ and the first term of $\mathcal{L}_\mathrm{cls}$ are computed on positive instants only. A bonafide utterance has no positive instants; on bonafide data, the CM is only trained by the second term of $\mathcal{L}_\mathrm{cls}$, where the focal loss at each instant becomes very small as soon as the score is moderately low. 

As a result, no loss tells the CM an utterance as a whole is bonafide and the CM still outputs proposals for bonafide inputs.
We therefore add an utterance-level head. The outputs of the six SGP blocks are concatenated along time and pooled by attentive statistics pooling~\cite{okabe2018attentive} into
a weighted mean and a weighted standard deviation.
A two-layer MLP with a hidden dimension of 256 and ReLU, followed by a sigmoid, maps the pooled vector to a probability $p_u$. 
The head is trained with class-balanced binary cross-entropy loss $\mathcal{L}_\mathrm{utt}$. During inference, an utterance judged bonafide by this head has no proposals.

\noindent\textbf{Frame-level prediction.} 
This auxiliary task predicts one spoof score per frame; it is used to compare with frame-level counterparts.
The head of this task takes the level 1's output of the SGP block, where one instant corresponds to one frame. It has the same structure as the instant classification head. It is trained with binary cross-entropy $\mathcal{L}_\mathrm{fr}$ against the quantized frame labels. The gradient is stopped at the input of the head, so the quantized labels do not affect the SGP blocks or the other heads.

Finally, the total loss is $\mathcal{L} = \mathcal{L}_\mathrm{cls} + \lambda_h \mathcal{L}_\mathrm{hub} + \mathcal{L}_\mathrm{IoU} +  \mathcal{L}_\mathrm{utt} + \mathcal{L}_\mathrm{fr}$, where $\lambda_h = 0.05$ brings $\mathcal{L}_\mathrm{hub}$ to the scale of $\mathcal{L}_\mathrm{IoU}$
at the start of training. 
Similar to other temporal action detection tasks~\cite{zhang2022actionformer, lin2019bmn, shi2023tridet}, we adopt Soft-NMS~\cite{bodla2017softnms} to merge near-duplicate proposals to a list of 100 proposals.

\section{Experimental settings}

\noindent\textbf{Training configuration.} All CMs are trained with the same recipe on
PartialSpoof~\cite{zhang2022partialspoof}, LlamaPartialSpoof~\cite{luong2025llamapartialspoof} and PartialEdit E1~\cite{zhang2025partialedit}: AdamW with learning rate $10^{-5}$ for XLS-R, $10^{-4}$ and weight decay 0.05 for the others, linear warm-up over 5 epochs and cosine decay over 30 epochs, batches of 8 whole utterances, early stopping with patience 3, and fp16 autocast for the front-end on one
48~GB RTX~A6000. The hyper-parameters were fixed on the PartialSpoof development set and not tuned per dataset: $\lambda_h = 0.05$, and a gated threshold of 0.2 on $p_u$.

\noindent\textbf{Evaluation configuration.} Every model is trained with three seeds and the mean is reported. We evaluate the CMs with four metrics: point-based EER, range-based EER~\cite{zhang2023rangeeer}, boundary errors (\emph{MAE s/e}) and matched fraction (Matched).
Point-based EER is computed at 20~ms resolution. For interval-based CMs, a frame takes the highest score among the proposals that cover it, and 0 if no proposal covers it. 
\emph{MAE s/e} is the mean absolute error of the predicted start and end times, in ms. \emph{Matched} is the percentage of reference spoofed intervals that a CM finds. A reference interval is matched if its best predicted interval has $\iou \ge 0.5$, and the MAE is computed between the reference and the highest-$\iou$ prediction. 
Frame-level CMs output no intervals, so consecutive frames with a normalized spoof score above 0.5 are stitched into one predicted interval.

\section{Results}

\noindent\textbf{Compared CMs.} Most published CMs for this task are frame-level classifiers, so we compare with several of them: BAM (boundary-aware attention over WavLM features)~\cite{zhong2024bam}, Split and Conquer (boundary detection followed by segment classification)\cite{rimon2026split}, and MRM (multi-resolution CM, the only CM reported with range-based EER)\cite{zhang2022partialspoof, zhang2023rangeeer}. The proposal-based CM closest to ours is CFPRF (coarse-to-fine proposal refinement on XLS-R)~\cite{wu2024cfprf}. 

\noindent\textbf{Our settings.} 1) \emph{ungated}, all proposals are included.
2) \emph{gated}, utterances judged bonafide by the utterance-level head get no proposals. 3) \emph{frame head}, use the frame-level head alone for inference. 4) \emph{max fusion}, the per-frame maximum of the gated proposal score and the frame head score.

\subsection{Comparison with published CMs on PartialSpoof}

\begin{table}[h]
\centering\footnotesize\setlength{\tabcolsep}{2pt}
\caption{Results on PartialSpoof. $^\dagger$: Our re-implementation.}
\label{tab:ps}
\begin{tabular*}{\columnwidth}{@{\extracolsep{\fill}}lcccc@{}}
\toprule
Model & Point EER & Range EER & MAE s / e & Matched \\
\midrule
MRM\cite{zhang2022partialspoof, zhang2023rangeeer} & 12.84 & 30.40 & -- & -- \\
MRM, re-impl.$^\dagger$ & 15.57 & 16.07 & 13.9 / 15.2 & 69.0 \\
CFPRF\cite{luong2025robust} & 7.61 & -- & -- & -- \\
SaC\cite{rimon2026split} & 6.55 & -- & -- & -- \\
BAM\cite{zhong2024bam} & 5.20 & -- & -- & -- \\
\midrule
Ours, w/o utt & 6.01 & 6.89 & 9.8 / 9.3 & \textbf{95.8} \\
Ours, ungated & 5.00 & 5.55 & \textbf{8.7 / 9.3} & 94.5 \\
Ours, gated & 5.00 & 5.56 & \textbf{8.7 / 9.3} & 94.3 \\
Ours, frame head & 5.89 & 6.36 & 14.2 / 17.9 & 78.9 \\
Ours, max fusion & \textbf{4.92} & \textbf{5.34} & 13.0 / 14.3 & 80.6 \\
\bottomrule
\end{tabular*}
\end{table}

Table~\ref{tab:ps} shows the results on PartialSpoof.
The row \emph{w/o utt} is the localizer trained without the utterance-level head; it measures the impact of adding this head.
Without the head, both point-based and range-based EERs rise, so the joint training itself helps localization. Gating during inference does not change the performance, but it has a practical value: it can filter out bonafide data and reduce post-processing time.

Range-based EER is higher than point-based EER in every row, because the former also counts the boundary errors within a frame. Compared to baselines, our CMs have lower point-based and range-based EERs in almost all settings. \textit{Max fusion} is the best; it beats using proposals alone and frame-level outputs alone, so the two outputs complement each other. We then compare \emph{MAE s/e} and \emph{Matched}. The regression localizers (ungated, gated) place both boundaries within 10~ms and match about 95\% of the reference intervals, whereas the frame-level CMs (frame head, max fusion) match only 78--80\% with errors above 13~ms. We attribute the difference to the training signal: the localizer learns from the exact reference, without quantization error.

\subsection{Generalization to LlamaPartialSpoof}

LlamaPartialSpoof (LPS) is used as an out-of-domain test: CMs trained on PartialSpoof are evaluated on it (zero-shot), or after fine-tuning on an LPS training subset (fine-tuned). Following the same setting of~\cite{luong2025robust}, we report point-based EER on its three groups of the test subset: \emph{full}, bonafide plus fully spoofed utterances; \emph{partial}, bonafide plus partially spoofed utterances; and \emph{both}, all utterances. 

\begin{table}[!t]
\centering\footnotesize\setlength{\tabcolsep}{1.5pt}
\caption{Point-based EER on LlamaPartialSpoof}
\label{tab:lps}
\begin{tabular*}{\columnwidth}{@{\extracolsep{\fill}}lccccccc@{}}
\toprule
& \multicolumn{3}{c}{Zero-shot} & & \multicolumn{3}{c}{Fine-tuned} \\
\cmidrule(r){2-4} \cmidrule(l){6-8}
Model & full & partial & both & & full & partial & both \\
\midrule
MRM\cite{zhang2022partialspoof, luong2025robust} & 40.98 & 32.76 & 44.54 & & 6.04 & 10.31 & 8.20 \\
CFPRF\cite{wu2024cfprf, luong2025robust} & 36.22 & 35.01 & 41.47 & & 2.89 & 8.72 & 5.96 \\
\midrule
Ours, ungated & 30.01 & 26.18 & 34.39 & & 3.23 & 1.76 & 2.85 \\
Ours, gated & 30.43 & 26.10 & 34.63 & & 3.04 & \textbf{1.71} & 2.78 \\
Ours, frame head & \textbf{28.75} & \textbf{24.47} & 33.84 & & 0.83 & 2.00 & \textbf{1.31} \\
Ours, max fusion & 29.57 & 24.60 & \textbf{33.79} & & 0.87 & 1.77 & 1.33 \\
Ours, utterance head & 37.03 & -- & -- & & \textbf{0.61} & -- & -- \\
\bottomrule
\end{tabular*}
\end{table}

Table~\ref{tab:lps} shows two trends that hold for all CMs used in these experiments. First, zero-shot evaluation does not work well: the point-based EER is between 24.47\% and 44.54\%, so a CM trained on PartialSpoof alone cannot be applied to LPS directly. Second, fine-tuning reduces the error largely. On \emph{both}, CFPRF drops from 41.47\% to 5.96\%, and our \textit{frame head} drops from 33.84\% to 1.31\%. 
Within these two trends, our CMs have the lower error in both zero-shot and fine-tuned conditions. In zero-shot, 
our settings have better results on every group, by 5.79--12.23\%, so training on exact boundaries generalizes better. 
After fine-tuning, ours achieve the best results: 1.71--2.00\% on \emph{partial} and 1.31--2.85\% on \emph{both}.  

One exception: \textit{ungated} performs worse than CFPRF on \emph{full}. After investigation, we found it is because the proposals do not always cover the whole spoofed utterance, indicating a task mismatch (utterance-level detection vs. partial spoof localization). Using the utterance-level head alone (row \textit{utterance head}) can reach 0.61\%, showing the robustness of the learned features.

\subsection{Revealing label misalignment issues}

\begin{table}[t]
\centering
\caption{Results on the PartialEdit E1 evaluation set,
scored against the realigned reference.
$^\dagger$: as reported in~\cite{zhang2025partialedit},
on the released labels and the authors' own evaluation set.}
\label{tab:pe}
\footnotesize\setlength{\tabcolsep}{2pt}
\begin{tabular*}{\columnwidth}{@{\extracolsep{\fill}}lcccc@{}}
\toprule
Model & Point EER & Range EER & MAE s / e & Matched \\
\midrule
\multicolumn{5}{@{}l}{\emph{Trained on the released labels}} \\
BAM$^\dagger$\cite{zhong2024bam, zhang2025partialedit} & 3.80 & -- & -- & -- \\
MRM, re-impl. & 7.43 & 7.51 & 10.7 / 127.3 & 89.4 \\
Ours, ungated & 7.23 & 7.65 & 25.2 / 82.7 & 99.0 \\
\midrule
\multicolumn{5}{@{}l}{\emph{Trained on the realigned labels}} \\
MRM, re-impl. & 0.53 & 0.54 & 10.0 / 8.4 & 99.0 \\
Ours, ungated & 1.05 & 0.78 & \textbf{7.2 / 5.8} & \textbf{99.5} \\
Ours, gated & 1.05 & 0.79 & \textbf{7.2 / 5.8} & 99.4 \\
Ours, frame head & \textbf{0.41} & 0.44 & 8.4 / 6.8 & 99.1 \\
Ours, max fusion & 0.44 & \textbf{0.43} & 9.6 / 7.4 & 99.3 \\
\bottomrule
\end{tabular*}
\end{table}

\begin{figure}[h]
\centering
\includegraphics[width=\columnwidth]{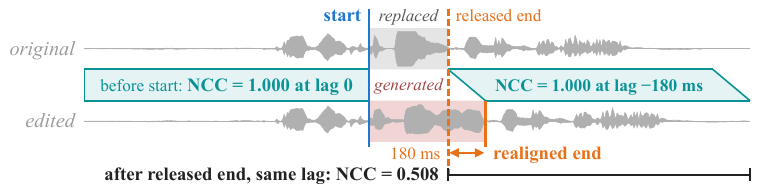}
\caption{Comparing the boundaries using cross-correlation.}
\label{fig:ncc}
\end{figure}

Table~\ref{tab:pe} shows the results on PartialEdit E1. Trained on the released labels, our CM has a higher point-based EER than BAM$^\dagger$. For a second reference, we include a re-implementation of MRM. The two CMs have one thing in common: the end error is several times the start error.
To check the labels, we compare each file that has a single spoofed interval with its VCTK source~\cite{yamagishi2019vctk} by normalized cross-correlation (NCC), as shown in Fig.~\ref{fig:ncc}. 
In terms of the starting point, the released time labels are quite accurate, and the edited and source files match ($\mathrm{NCC} > 0.99$) for 92\% of the files. 
As for the ending time label, however, the ratio of matched files is only 37\%.

Therefore, we realigned all single-interval files 
and created a `realigned' subset from PartialEdit E1\footnote{\url{https://huggingface.co/datasets/DeepFense/partialedit-corrected}}. 
Trained on this subset, both CMs drop from above 7\% to about 1\% point-based EER or below, and our CMs have the best \emph{MAE s/e} and \emph{Matched}.
The reference MRM model also benefits from using the realigned training data.

\section{Conclusion}

We proposed a single-stage CM that treats spoof localization as boundary regression, predicting the start time and end time from the exact reference without quantization. It achieves the lowest point-based and range-based EER on PartialSpoof, and generalizes better to LlamaPartialSpoof. Boundary error analysis led us to realign PartialEdit.

\section{Compliance with Ethical Standards}
This study was conducted using speech corpora made available in open access (PartialSpoof, LlamaPartialSpoof, PartialEdit and VCTK). Ethical approval was not required as confirmed by the licenses attached to the open-access data. 

\bibliographystyle{IEEEbib}
\bibliography{strings,refs}

\clearpage
\appendix
\section{Interpreting the evaluation metrics of partial spoof localization}
\label{app:rangeeer}
This appendix recaps the computation of equal error rates (EERs) for conventional binary classification and extends it to the evaluation metrics for partial-spoof detection \& localization. The contents are listed below:
\begin{enumerate}
    \item On binary classification task (\S~\ref{sec:binary}): as pointed out by~\cite{brummer2021out}, EER is the maximum of the Bayes decision cost w.r.t the prior, given that the scores are perfectly calibrated.
    \item On partial spoofing (\S~\ref{sec:partialspoof}): 
    \begin{itemize}
        \item Segment-EER (\S~\ref{sec:sub:segeer}) can be interpreted in the same manner as the EER for conventional binary classification, assuming that the segment scores are independently identically distributed. 
        \item Range EER (\S~\ref{sec:sub:rangeeer}) assumes that the score $r(t)$ as a function of time. Interpretation of range EER in a similar manner to the conventional EER requires strong assumptions about the score distribution.
    \end{itemize}
\end{enumerate}

\subsection{Binary classification}
\label{sec:binary}

\subsubsection{Optimal decision}
\label{sec:sub:decision}
The user of a binary classifier needs to guess or make a decision whether the input waveform $\evidence\in\mathcal{X}$ is real or fake. The decision is binary, and the target label can be $\mathcal{Y}=\{\symbolhp{}_\classtar{}, \symbolhp{}_\classnon{}\}$, where
\begin{align*} 
\symbolhp{}_\classtar{}:  & \, \evidence \text{ is REAL}, \\
\symbolhp{}_\classnon{}: & \,\evidence \text{ is FAKE}.
\end{align*} 
The user action $a$ can be from $\mathcal{A} = \{\acty, \actn\}$:
\begin{align*}
\acty: &\, \text{user claims that } \evidence  \text{ is REAL },\\
\actn: &\, \text{user claims that } \evidence  \text{ is FAKE }.
\end{align*}

The user uses a detection system when making the decision, the $\evidence$ is consumed and converted into a score $\score\in\mathbb{R}$, where the value indicates how likely that $\evidence$ is REAL. Intuitively, the user should take $a=\acty$ if $\score$ is high enough, indicating a higher chance of being real. 

The user may take an incorrect action: a real sample ($H_\classtar{}$) may be falsely rejected (\emph{miss}), while a fake sample ($H_\classnon{}$) may be falsely accepted (\emph{false acceptance}). Incorrect actions can be assigned decision costs.
Given an action $a\in\mathcal{A}$ and the ground-truth label $y\in\mathcal{Y}$, we use $C(a, y)\geq0$ to denote the decision-cost function $c: \mathcal{A}\times{\mathcal{Y}}\rightarrow\mathbb{R}^{+}$. An example is given in Table~\ref{tab:app_cost}. 

Following the practice of Bayes decision~\cite{dudaPattern2001}, the user's conditional risk $R(a | \score)$ when taking an action $a$ given the input $\score$ is computed as
\begin{equation}
R(a | \score) = \sum_{y\in\mathcal{Y}}C(a, y) P(y | \score), 
\end{equation}
where $P(y | \score)$ is the posterior probability of class $y$ when observing $\score$. 
By plugging in the possible values of $C(a, y)$ listed in Table~\ref{tab:app_cost}, we can compute the conditional risks when taking the two possible actions
\begin{align}
R(\acty | \score) &= \costfanon{} P(\symbolhp{}_\classnon{} | \score), \label{eq:app_cost1} \\ 
R(\actn | \score) &= \costmiss{} P(\symbolhp{}_\classtar{} | \score). \label{eq:app_cost2}
\end{align}
The Bayes decision rule tells the user to select the action $a$ with a smaller risk~\cite{dudaPattern2001}, i.e.,
\begin{equation}
{a} = 
\begin{cases}
\acty, \quad \text{if } R(\actn | \score) > R(\acty | \score) \\
\actn, \quad \text{if } R(\actn | \score)  \leq R(\acty | \score) \\
\end{cases}.
\end{equation}
\begin{table}[!t]
\caption{Values of decision cost $C(a, y)$. Subscript $_\text{fa}$ and $_\text{fr}$ denote false $\acty$ and $\actn$, respectively.}
\vspace{-5mm}
\begin{center}
\begin{tabular}{rcll}
\toprule
 && Action = $\acty$ & Action = $\actn$\\
\midrule
True class = $\classtar{}$  && 0 & $\costmiss$ \\
True class = $\classnon{}$ && $\costfanon$ & 0\\
\bottomrule
\end{tabular}
\end{center}
\label{tab:app_cost}
\end{table}%
With Eqs.(\ref{eq:app_cost1}) and (\ref{eq:app_cost2}), the policy above can be re-written as 
\begin{equation}
\costmiss{} P(\symbolhp{}_\classtar{} | \score) \underset{\actn}{\overset{\acty}{\gtrless}}  \costfanon{} P(\symbolhp{}_\classnon{} | \score).
\end{equation}
From the Bayes' formula, the user knows that $P(\symbolhp{}_\classtar{} | \score) = p(\score | \symbolhp{}_\classtar{}) \pi_{\classtar{}}/p(s)$ and $P(\symbolhp{}_\classnon{} | \score)  
= p(\score | \symbolhp{}_\classnon{} ) \pi_{\classnon{}}/p(s)$,  
where $\pi_{\classtar{}}\triangleq P(H_\classtar{}) = 1-\pi_{\classnon{}}$ is the prior probability of $H_\classtar$. The user can then further rewrite the policy in a log-ratio form as
\begin{equation}
 \log\frac{p( \score | \symbolhp{}_\classtar{})}{p( \score | \symbolhp{}_\classnon{} )} \underset{\actn}{\overset{\acty}{\gtrless}}  \log\Big(\frac{(1-\pi_\classtar)}{\pi_\classtar} \frac{\costfanon{}}{\costmiss{}}\Big).
\label{eq:app_policy_gen}
\end{equation}

The term on the left-hand side of Eq.(\ref{eq:app_policy_gen}) is the LLR of the observed $\score$, $\mathsf{LLR}(\score) \triangleq \log\frac{p( \score | \symbolhp{}_\classtar{})}{p( \score | \symbolhp{}_\classnon{} )}$. This is the only term that depends on the input speech data (via $\score$). The logit $\log\frac{\pi_\classtar}{(1-\pi_\classtar)}$ and the term for decision costs are set \emph{a prior} by the user. 
In other words, when the user's detection system is launched, the prior and decision costs are configured by the user. Given an input $\evidence$, the detector produces $\score$, based on which the user makes a decision following Eq.(\ref{eq:app_policy_gen}). 

This is the optimal decision policy for the user.

\subsubsection{Interpretation of EER}
\noindent\textbf{When user has ideal LLRs and their distributions.}
Let $r\triangleq \mathsf{LLR}(s)\triangleq\log\frac{p( \score | \symbolhp{}_\classtar{})}{p( \score | \symbolhp{}_\classnon{} )}$ be a short-hand notation of the LLR value corresponding to $\score$. 
With the decision costs plugged in, the overall risk $R(\pi_\classtar, {\costfanon}, {\costmiss})$ when the user follows the optimal Bayes decision policy in Eq.(\ref{eq:app_policy_gen}) becomes:
\begin{align}
    &R(\pi_\classtar, {\costfanon}, {\costmiss}) 
    = \sum_{a\in\mathcal{A}}\int R(a | r) p(r)dr \\ 
    = &\int\sum_{a\in\mathcal{A}} \sum_{y\in\mathcal{Y}}C(a, y) P(y | r)p(r)dr\\
    =&\int_{\tau_\text{Bayes}}^{+\infty} {\costfanon}P(H_\classnon | r)p(r)d r + \int_{-\infty}^{\tau_\text{Bayes}} {\costmiss}P(H_\classtar | r)p(r)dr \\
    =&{\costfanon}(1-\pi_\classtar) \int_{\tau_\text{Bayes}}^{+\infty} p( r | H_\classnon) dr  +  {\costmiss}\pi_\classtar \int_{-\infty}^{\tau_\text{Bayes}}p(r |  H_\classtar )dr \label{eq:overallrisk_llr}\\
    =&{\costfanon}(1-\pi_\classtar) p_{\text{fa}}(\tau_\text{Bayes}) +  {\costmiss}\pi_\classtar p_{\text{miss}}(\tau_\text{Bayes}),
    \label{eq:bayesrisk}
\end{align}
where $p_{\text{fa}}(\tau_\text{Bayes})$ and $p_{\text{miss}}(\tau_\text{Bayes})$ correspond to the probabilities of false $\acty$ and $\actn$, respectively,
\begin{align}
p_{\text{fa}}(\tau_\text{Bayes})&\triangleq\int_{\tau_\text{Bayes}}^{+\infty} p( r | H_\classnon) dr, \label{eq:pfa:llr}\\
p_\text{miss}(\tau_\text{Bayes})&\triangleq \int_{-\infty}^{\tau_\text{Bayes}}p(r |  H_\classtar )dr,\label{eq:pmiss:llr}
\end{align}
and where $\tau_\text{Bayes}=\log\Big({(1-\pi_\classtar)\costfanon{}}/({\pi_\classtar\costmiss{}})\Big)$ is the optimal Bayes decision threshold decided by the user's asserted prior and decision costs.\footnote{Row (\ref{eq:overallrisk_llr}) is obtained because  $P(H_\classnon | r)p(r)=(1-\pi_\classtar) p( r | H_\classnon)$ and $P(H_\classtar | r)p(r)=\pi_\classtar p( r | H_\classtar)$. Furthermore, the prior and decision costs are independent of $r$ and moved outside of the integration.}

The notion that the overall risk $R$ is a function of $\{\pi_\classtar, {\costfanon}, {\costmiss}\}$ reflects the fact that, given the $r$ and its distributions, the user's decision risk is only affected by these three user-asserted parameters. If the user alters them, the decision threshold $\tau_\text{Bayes}$ and the corresponding overall risk are changed automatically. 

It has been shown that~\cite[\S~2]{brummer2021out}, given $C_\text{fa}^\classnon={\costmiss}=1$, EER is the maximum overall risk when the user varies $\pi_\classtar$. 
\begin{align}
    \text{EER} &= \max_{\pi_\classtar} R(\pi_\classtar, {\costfanon}=1, {\costmiss}=1) \label{eq:eerdef}\\
    &= \max_{\pi_\classtar} \Big((1-\pi_\classtar) p_{\text{fa}}(\tau(\pi_\classtar)) +  \pi_\classtar p_{\text{miss}}(\tau(\pi_\classtar))\Big),
\end{align}
where the threshold $\tau(\pi_\classtar)=\log\frac{1-\pi_\classtar}{\pi_\classtar}$ is a function of $\pi_\classtar$. The $\pi_\classtar^*$ that yields the maximum corresponds to the threshold $\tau_\text{EER}=\log\frac{1-\pi_{\classtar}^*}{\pi_{\classtar}^*}$, which also satisfies $p_{\text{fa}}(\tau_\text{EER}) = p_\text{miss}(\tau_\text{EER})$,
hence the name equal error rate.

Here, we write down the proof again. Let $\mathrm{R}(\pi_\classtar) \triangleq R(\pi_\classtar, {\costfanon}=1, {\costmiss}=1)$ be a shorthand of the decision risk. By taking the first-order derivative w.r.t. $\pi_\classtar$, we have
\begin{align}
    \frac{d \mathrm{R}(\pi_\classtar)}{d\pi_\classtar} = & -p_{\text{fa}}(\tau(\pi_\classtar)) + (1-\pi_\classtar)  \frac{d p_{\text{fa}}(\tau(\pi_\classtar))}{d \pi_\classtar} \\ & + 
    p_{\text{miss}}(\tau(\pi_\classtar)) + \pi_\classtar\frac{d p_{\text{miss}}(\tau(\pi_\classtar))}{d\pi_\classtar},
    \label{eq:deri1st}
\end{align} 
where the derivatives of the two error rates are
\begin{align}
    \frac{d p_{\text{fa}}(\tau(\pi_\classtar))}{d \pi_\classtar}  =& -p(\tau(\pi_\classtar)|H_\classnon{})\frac{d\tau(\pi_\classtar)}{d\pi_\classtar}\\
    \frac{d p_{\text{miss}}(\tau(\pi_\classtar))}{d \pi_\classtar}  =& p(\tau(\pi_\classtar)|H_\classtar{})\frac{d\tau(\pi_\classtar)}{d\pi_\classtar},
\end{align}
and where
\begin{equation}
    \frac{d\tau(\pi_\classtar)}{d\pi_\classtar} = \frac{-1}{(1-\pi_\classtar)\pi_\classtar}.
\end{equation}
Because of the property that the LLR of LLR is LLR~\cite{brummer2021out}, i.e., $r = \log\frac{p( r | \symbolhp{}_\classtar{})}{p( r | \symbolhp{}_\classnon{} )}$, 
at any decision threshold $r=\tau(\pi_\classtar)$, we know that 
\begin{equation}
    p( \tau(\pi_\classtar) | \symbolhp{}_\classtar{}) = \exp(\tau(\pi_\classtar))p( \tau(\pi_\classtar) | \symbolhp{}_\classnon{} ) = \frac{1-\pi_\classtar}{\pi_\classtar}p( \tau(\pi_\classtar) | \symbolhp{}_\classnon{} ). 
    \label{eq:llrisllr}
\end{equation}
With the above four equations, it can be shown that
\begin{align}
&(1-\pi_\classtar)  \frac{d p_{\text{fa}}(\tau(\pi_\classtar))}{d \pi_\classtar}  + \pi_\classtar\frac{d p_{\text{miss}}(\tau(\pi_\classtar))}{d\pi_\classtar} \\
=&     -(1-\pi_\classtar) p(\tau(\pi_\classtar)|H_\classnon{})\frac{d\tau(\pi_\classtar)}{d\pi_\classtar} + \pi_\classtar p(\tau(\pi_\classtar)|H_\classtar{})\frac{d\tau(\pi_\classtar)}{d\pi_\classtar} \\
=& \Big(    -(1-\pi_\classtar)  + \pi_\classtar \frac{(1-\pi_\classtar)}{\pi_\classtar} \Big) p(\tau(\pi_\classtar)|H_\classnon{})\frac{d\tau(\pi_\classtar)}{d\pi_\classtar}\\
=&0
\end{align}
Hence, the 2nd and 4th terms in Eq.(\ref{eq:deri1st}) cancel out, and Eq.(\ref{eq:deri1st}) becomes
\begin{equation}
    \frac{d \mathrm{R}(\pi_\classtar)}{d\pi_\classtar} = p_{\text{miss}}(\tau(\pi_\classtar)) - p_{\text{fa}}(\tau(\pi_\classtar)).
\end{equation}
Taking the 2nd order derivative, we have
\begin{equation}
    \frac{d^2 \mathrm{R}(\pi_\classtar)}{d\pi_\classtar^2} = \Big(p(\tau(\pi_\classtar) | H_\classtar) + p(\tau(\pi_\classtar) | H_\classnon)\Big)\frac{-1}{(1-\pi_\classtar)\pi_\classtar} < 0.
\end{equation}
Hence, $\mathrm{R}(\pi_\classtar^*)$ has a global maximum when the variable $\pi_\classtar^*$ satisfies that $\frac{d \mathrm{R}(\pi_\classtar)}{d\pi_\classtar} |_{\pi_\classtar = \pi_\classtar*} = p_{\text{miss}}(\tau(\pi_\classtar^*)) - p_{\text{fa}}(\tau(\pi_\classtar^*))=0$.
In other words, the decision threshold $\tau(\pi_\classtar^*)=\log\frac{1-\pi_{\classtar}^*}{\pi_{\classtar}^*}$ makes the two error rates equal.

In short, EER is the maximum risk w.r.t to the user's asserted prior $\pi_\classtar$. The above interpretation requires that 
\begin{enumerate}
    \item The decision is based on perfect LLRs $r$, rather than uncalibrated scores.
    \item The decision is based on the Bayes decision policy (\ref{eq:app_policy_gen}). 
    \item The risk $\mathrm{R}$ assumes fixed and equal decision costs to the two types of decision errors. 
\end{enumerate}

\begin{figure*}[!t]
    \centering
    \includegraphics[width=0.8\linewidth, trim=0 320 0 0, clip]{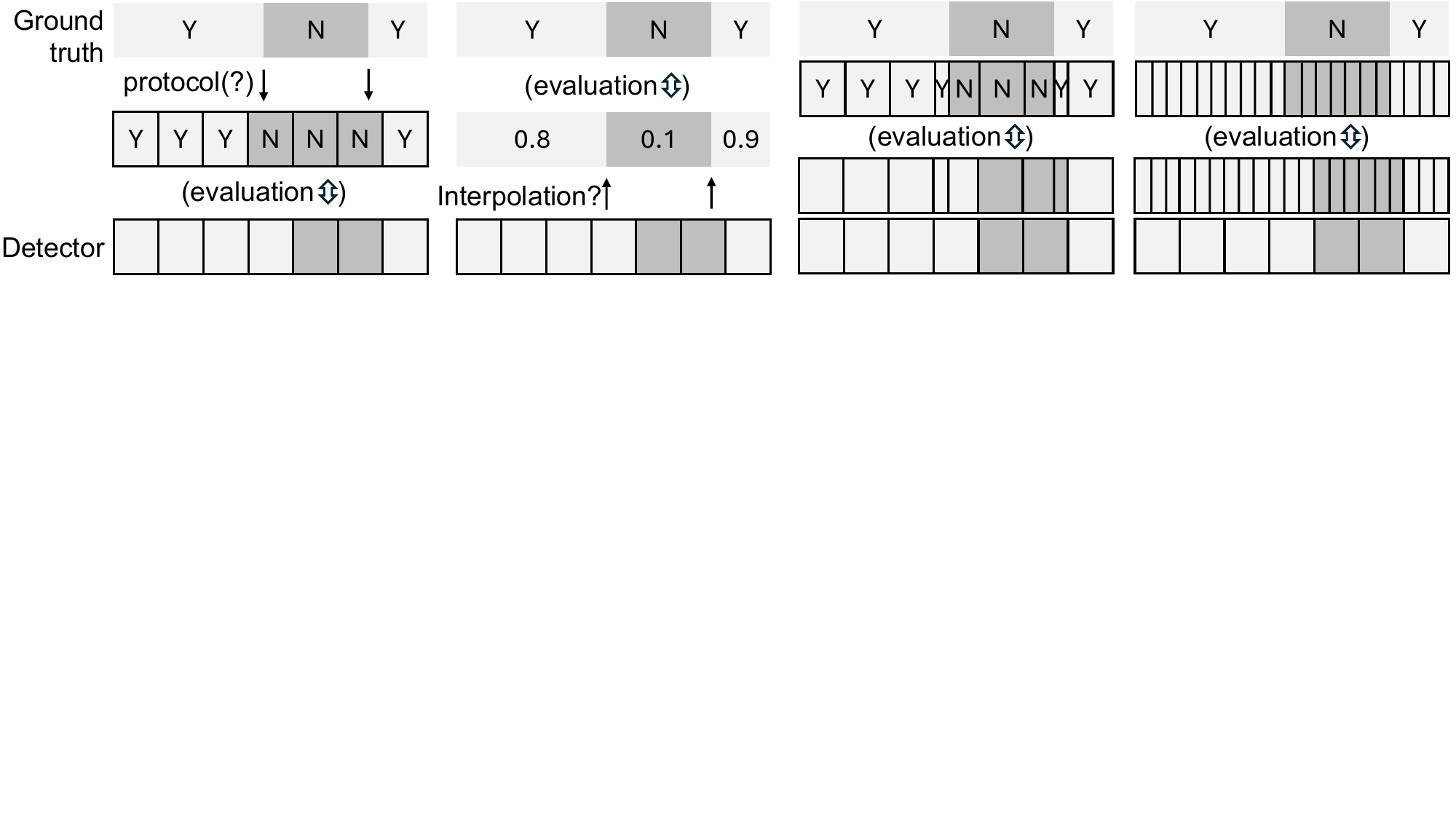}
    \caption{Three ways of solving alignment issue between ground truth and model outputs for evaluation of partial spoofing detection.}
    \label{fig:alignment}
\end{figure*}

\noindent\textbf{When user has ideal LLRs but not their distributions.}

In reality, the user does not know the true $p(r|H_\classtar)$ and $p(r|H_\classnon)$ and cannot compute the probabilities of false $\acty$ and $\actn$ by integration. The user has to estimate the error probabilities on a test set by counting. 

Let $\mathcal{D}_{r,\classtar}=\{r_i | H_\classtar\}$ and $\mathcal{D}_{r,\classnon}=\{r_i | H_\classnon\}$ be the LLRs of the $H_\classtar$ and $H_\classnon$ samples in a test set. 
Let a general threshold be $\tau$, which may be different from the optimal $\tau_\text{Bayes}$. The error rates are estimated by
\begin{equation}
    \widehat{P}_\text{fa}(\tau) = \frac{1}{|\mathcal{D}_{r,\classnon}|}\sum_{r_i\in\mathcal{D}_{r,\classnon}}\mathrm{1}(r_i>\tau),
    \label{eq:pfa_c:llr}
\end{equation}
\begin{equation}
    \widehat{P}_\text{miss}(\tau) = \frac{1}{|\mathcal{D}_{r,\classtar}|}\sum_{r_i\in\mathcal{D}_{r,\classtar}}\mathrm{1}(r_i<\tau),
    \label{eq:pmiss_c:llr}
\end{equation}
where $\mathrm{1}(\cdot)$ is an indicator function.
Plugging in the estimated error rates, the user gets the estimated overall risk
\begin{equation}
    \widehat{R}(\pi_\classtar, {\costfanon}, {\costmiss},\tau) = {\costfanon}(1-\pi_\classtar) \widehat{P}_{\text{fa}}(\tau) +  {\costmiss}\pi_\classtar \widehat{P}_{\text{miss}}(\tau).
\end{equation}
Note that this is the decision cost function (DCF), for example, used in NIST SRE~\cite[Eq.(2)]{nistNIST2016}. 
If the threshold is set according to Bayes decision rule $\tau_\text{Bayes}=\log\Big({(1-\pi_\classtar)\costfanon{}}/({\pi_\classtar\costmiss{}})\Big)$, the resulting DCF is an approximation of Bayes risk in Eq.(\ref{eq:bayesrisk}), where the two types of error rates area are approximated by counting rather than integration. 

Hence, when $C_\text{fa}^\classnon={\costmiss}=1$,  the property of EER holds.
However, since $P_\text{fa}(\tau)$ and $P_\text{miss}(\tau)$ are not continuous functions of $\tau$ anymore, it may be infeasible to find a $\tau_\text{EER}$ so that $P_{\text{fa}}(\tau_\text{EER})$ is exactly equal to $P_\text{miss}(\tau_\text{EER})$. As an approximation, the threshold can be set to \begin{equation}
    \hat{\tau}_\text{EER} \approx \arg\min_{\tau} | \widehat{P}_\text{miss}(\tau) - \widehat{P}_\text{fa}(\tau) |,
\end{equation}
and the EER can be approximated as
\begin{equation}
    \widehat{\text{EER}}\approx\Big(\widehat{P}_\text{miss}(\hat{\tau}_\text{EER}) + \widehat{P}_\text{fa}(\hat{\tau}_\text{EER})\Big)/2.
    \label{eq:eer:approximate}
\end{equation}

The same requirements in the previous subsection are satisfied (or assumed to be true). In the rest of the paper, we also assume that the detector produces perfect LLRs. 

\subsection{Detection for Partial Spoof}
\label{sec:partialspoof}

\subsubsection{Alignment between ground-truth and detector output}
In a partial spoof scenario, an utterance is a concatenation of at least one real and one fake segment. The first task is to make a decision on each segment. 

The setup is different from the utterance-level binary classification in the previous section. A major reason is that segments have arbitrary length, while a detector may only produce a score at a uniform time interval. This leads to an alignment issue for evaluation.

Before discussing different ways of handling the alignment issue, we need to agree on the principle of evaluation: for evaluation that need both ground truth and model output, the ground truth should be the SAME regardless of model behavior. 

There may be three ways (shown in Figure~\ref{fig:alignment}) to handle that:
\begin{enumerate}
    \item Converting the ground-truth labels so that a ground-truth label can be assigned at the same uniform interval. 
    \item Converting the detector's output scores to a score for each of the segment defined by the ground-truth label.      
    \item Converting both the ground-truth labels and the model outputs so that a boundary is inserted whenever it hits the boundary of the ground truth OR that of a uniform segment.
    \item Converting both the ground-truth labels and the model outputs by splitting both with an interval $\Delta>0$. 
\end{enumerate}

We can argue that the first strategy is suboptimal because it violates the principle we agree on.  A consequence of violating the principle is that we need additional `agreement' on how to convert ground-truth labels when there is ambiguity across the fixed-length interval (e.g., the left- and right-most `N' in middle left of Fig.~\ref{fig:alignment}). \footnote{The agreement can be somehow reached if an evaluation set comes with the pre-defined sets of ground-truth labels~\cite{zhang2022partialspoof}.} 

Furthermore, the conversion of ground-truth label is affected by the detector's capability, (which should NOT be). If two detectors predict a score per 8ms and 15ms, respectively, their `ground-truth' labels are different by nature. Evaluation results using different ground truth are not comparable.

The 3rd strategy is also against the principle. The converted ground-truth depends on the behavior of the detector. 

The 2nd strategy is better because the ground-truth never changes. What changes is the model's output --- if the detector produces a score every 8ms, it has to include a post-processing module so that a score can be obtained for each of the segment defined by the ground-truth label. It is the detector's responsibility to produce a score in the format required by the evaluation metric. 

The 4th strategy uses an interval $\Delta$ independent of the detector's design. If $\Delta$ is small enough, we can get aligned ground-truth and model outputs. 

\subsubsection{Problem setup}

Regardless of the choice, we can wrap the definition of the task as this: the user needs to guess or make a decision whether the input segment, which spans from time $t_s$ to $t_e$ of the input utterance $\evidence\in\mathcal{X}$, is real or fake. The decision is binary, and the target label can be $\mathcal{Y}=\{\symbolhp{}_\classtar{}, \symbolhp{}_\classnon{}\}$, where
\begin{align*} 
\symbolhp{}_\classtar{}:  & \, \evidence(t_s, t_e) \text{ is REAL}, \\
\symbolhp{}_\classnon{}: & \, \evidence(t_s, t_e) \text{ is FAKE}.
\end{align*} 
The user action $a$ can be from $\mathcal{A} = \{\acty, \actn\}$:
\begin{align*}
\acty: &\, \text{user claims that } \evidence(t_s, t_e)  \text{ is REAL },\\
\actn: &\, \text{user claims that } \evidence(t_s, t_e)  \text{ is FAKE }.
\end{align*}

\subsubsection{Segment EER}
\label{sec:sub:segeer}
Originally segment EER~\cite{zhang2023rangeeer} assumes the first alignment strategy. Accordingly, the $(t_s, t_e)$ for each sample is defined by the detector's `time resolution': the time interval when the detector produces a score. The value of $t_e - t_s$ is equal to the `time resolution. 
The Bayes decision, decision risk, and the EER can be derived in the same manner as the utterance-level classification in Section~\ref{sec:binary}. The time information $(t_s, t_e)$ does not affect the error cost, and all the samples are assumed to be independent from each other. 

\subsubsection{Range EER}
\label{sec:sub:rangeeer}
\noindent\textbf{False rejection and false acceptance rate.}
The reason to switch to range EER is argued to be an ambiguity issue when converting the ground-truth labels~\cite{zhang2023rangeeer}. The mitigating strategy is the range EER. Here is one detailed explanation of the idea. 

Range-EER assumes that the 4th alignment strategy. The problem setup is similar to the segment EER, but the data to be scored becomes $\evidence(t, t+\Delta)$---we need a detector can produce a score for $\evidence(t, t+\Delta), \forall t, \forall\Delta$\footnote{To get the score from a detector that only produces a score per $N$ ms, the task of producing score at arbitrary $t$ is similar to digital-to-analog conversion \url{https://en.wikipedia.org/wiki/Digital-to-analog_converter}.}.

Without loss of generality, let us assume that one utterance has only one fake segment $(t_s, t_e)$. 
The number of (sub-) segments is $\frac{(t_e - t_s)}{\Delta}$.\footnote{The number of sub-segments must be an integer, but we drop the ceiling or flooring operator to avoid notation clutter.}
Let the score $r(t)$ be a function of time $t$. 
Let the decision threshold be $\tau$ and the LLR score of the $i$-th sub-segment be $r(t_s + i\Delta)$; we can then compute the false acceptance rate as
\begin{align}
    \widehat{P}_\text{fa}(\tau) &= \lim_{\Delta \to 0} \frac{1}{(t_e - t_s)/\Delta} \sum_{i=1}^{\frac{t_e - t_s}{\Delta}} \mathrm{1}(r(t_s + i\Delta)>\tau)  \\
    &= \lim_{\Delta \to 0} \frac{1}{(t_e - t_s)} \sum_{i=1}^{\frac{t_e - t_s}{\Delta}} \mathrm{1}(r(t_s + i\Delta)>\tau) \Delta \\
    &= \frac{1}{t_e - t_s} \int_{t_s}^{t_e} \mathrm{1}(r(t) > \tau) \, dt \label{eq:pfa_range:llr_continuous}
\end{align}
where $\mathrm1(\cdot)$ is an indicator function.
By taking the limit that $\Delta\rightarrow 0$, the integration is obtained based on the definition of  Riemann sum\footnote{\url{https://en.wikipedia.org/wiki/Riemann_sum}}. The visual intuition is that $\widehat{P}_\text{fa}(\tau)$ corresponds to the ratio of the interval wherein the score $r(t)$ within the interval is above the threshold $\tau$. 

Intuitively, given an evaluation set, we can concatenate all the (ground-truth) fake segments as well as their score functions into a single $r(t)$. The definition of false acceptance is simply 
\begin{equation}
    \widehat{P}_\text{fa}(\tau) = \frac{1}{T_\classnon} \int_{0}^{T_\classnon} \mathrm{1}(r(t) > \tau) \, dt,
    \label{eq:pfa_range}
\end{equation}
where $T_\classnon$ is the length of the concatenated fake segment. 
Similarly, we can define the false rejection rate as 
\begin{equation}
    \widehat{P}_\text{miss}(\tau) = \frac{1}{T_\classtar} \int_{0}^{T_\classtar} \mathrm{1}(r(t) < \tau) \, dt
    \label{eq:pmiss_range}
\end{equation}

The above definitions are slightly different from those in the original paper in two aspects: 1) the fake  (or real) segments are concatenated into a single trunk, which is to simplify the notation and does not affect the result\footnote{The original paper compares each score with each ground-truth segment, via the two summation operations. This is just to find the overlapped regions.}; 2) the definition takes the limit of $\Delta\rightarrow0$ and turns the summation into integration. Of course, in practice, we have to implement it via summation with a small $\Delta$.

\noindent\textbf{Bayes decision?}
Range EER simply finds the threshold $\tau_\text{EER}$ so that $\widehat{P}_\text{miss}(\tau_\text{EER}) = \widehat{P}_\text{fa}(\tau_\text{EER})$. This definition is not refutable. However, the question is whether this operation point $\tau_\text{EER}$ corresponds to the maximum decision risk (w.r.t to the prior) when taking a Bayes decision. 

Assuming uncalibrated scores $s(t)$ and decision making for $t$ independently, we can follow the optimal Bayes decision in \S~\ref{sec:sub:decision}  write the policy as 
\begin{equation}
 \log\frac{p( \score(t) | \symbolhp{}_\classtar{})}{p( \score(t) | \symbolhp{}_\classnon{} )} \underset{\actn}{\overset{\acty}{\gtrless}}  \log\Big(\frac{(1-\pi_\classtar)}{\pi_\classtar} \frac{\costfanon{}}{\costmiss{}}\Big),
\label{eq:app_policy_gen_range}
\end{equation}
where $\tau_\text{Bayes}\triangleq\log\Big(\frac{(1-\pi_\classtar)}{\pi_\classtar} \frac{\costfanon{}}{\costmiss{}}\Big)$ is the decision threshold defined by the prior and costs. 
Then we can let $r(t)\triangleq \mathsf{LLR}(s(t))\triangleq\log\frac{p( \score(t) | \symbolhp{}_\classtar{})}{p( \score(t) | \symbolhp{}_\classnon{} )}$ and write the decision risk (tentatively) as 
\begin{align}
    &R(\pi_\classtar, {\costfanon}, {\costmiss}) 
    \\ 
    =&{\costfanon}(1-\pi_\classtar) \frac{1}{T_\classnon}\int_0^{T_\classnon}\int_{\tau_\text{Bayes}}^{+\infty} p( r(t) | H_\classnon) drdt \\
     &+  {\costmiss}\pi_\classtar \frac{1}{T_\classtar}\int_0^{T_\classtar}\int_{-\infty}^{\tau_\text{Bayes}}p(r(t) |  H_\classtar )drdt 
    \label{eq:bayesrisk:range}
\end{align}
The difference from the case of conventional binary classification (Eq.(\ref{eq:bayesrisk})) is that the score is time-variant, and $r(t)$ can be treated as a random process.\footnote{Currently, there is no theory to explain why the risk should be defined in this way. It is just based on intuition that risk over time should be integrated.}

To simplify this double integration, we assume that $r(t)$ is \emph{stationary}: $p(r(t)|H) \equiv p(r|H)$. Consequently, the integrals over $t$ is averaged out:
\begin{align}
     &R(\pi_\classtar, {\costfanon}, {\costmiss}) \\
    = & {\costfanon}(1-\pi_\classtar) \int_{\tau_\text{Bayes}}^{+\infty} p( r | H_\classnon) dr \nonumber + {\costmiss}\pi_\classtar \int_{-\infty}^{\tau_\text{Bayes}}p(r | H_\classtar )dr.
    \label{eq:bayesrisk:range_stationary}
\end{align}
Under the stationary assumption, the risk becomes in the same format as the case of conventional binary classification (Eq.(\ref{eq:bayesrisk})). Similar to the proof in the previous section, by setting the derivative to zero we can obtain the maximum decision risk at the Bayes threshold $\tau_\text{EER} = \tau(\pi_\classtar^*)$ where:
\begin{equation}
\int_{\tau_\text{EER}}^{+\infty} p( r | H_\classnon) dr
    = \int_{-\infty}^{\tau_\text{EER}}p(r | H_\classtar )dr.
\end{equation}

The next step is to link above integration with the definitions of $\widehat{P}_\text{fa}(\tau)$ and $\widehat{P}_\text{miss}(\tau)$. We must invoke the property of \emph{ergodicity}: the time-average of a single realization of a stochastic process (over a sufficiently long period) converges to the ensemble average across all possible realizations. Assuming that $r(t)$ is ergodic, as the total dataset duration grows large ($T_\classnon \to \infty$), the empirical time-averaged false acceptance rate converges almost surely to the theoretical ensemble probability:
\begin{align}
    \lim_{T_\classnon \to \infty} \widehat{P}_\text{fa}(\tau) &= \lim_{T_\classnon \to \infty} \frac{1}{T_\classnon} \int_{0}^{T_\classnon} \mathrm{1}(r(t) > \tau) \, dt \\
    &= \mathbb{E}_{r}\big[ \mathrm{1}(r > \tau) \big| H_\classnon \big] \\
    &= \int_{\tau}^{+\infty} p(r | H_\classnon) \, dr.
\end{align}
Similarly, we have 
\begin{align}
    \lim_{T_\classtar \to \infty} \widehat{P}_\text{miss}(\tau) &= \lim_{T_\classtar \to \infty} \frac{1}{T_\classtar} \int_{0}^{T_\classtar} \mathrm{1}(r(t) < \tau) \, dt \\
    &= \mathbb{E}_{r}\big[ \mathrm{1}(r < \tau) \big| H_\classtar \big] \\
    &= \int^{\tau}_{-\infty} p(r | H_\classtar) \, dr.
\end{align}
Hence, the threshold $\tau_\text{EER}$ should satisfy that
\begin{align}
    \lim_{T_\classtar\to\infty}\widehat{P}_\text{miss}(\tau_\text{EER}) = &\int_{\tau_\text{EER}}^{+\infty} p( r | H_\classnon) dr
     = \int_{-\infty}^{\tau_\text{EER}}p(r | H_\classtar )dr \\ 
    &\quad = \lim_{T_\classnon\to\infty}\widehat{P}_\text{fa}(\tau_\text{EER}).
\end{align}
Hence, on a large evaluation set, finding the range-EER is likely to have the same interpretation as the conventional EER: the maximum risk w.r.t the Bayes decision threshold (or the prior). 

However, the above derivation requires
\begin{enumerate}
    \item Stationary assumption: $p(r(t)|H) \equiv p(r|H)$.
    \item Ergodic assumption: the time-average of a single realization of a stochastic process ($r(t)$) converges to the ensemble average across all possible realizations $\mathbb{E}_{R(t)}$.
    \item LLR $r(t)$ is used for optimal Bayes decision making and decision costs $C_\text{fa}=C_\text{fr}=1$.
\end{enumerate}
Whether the assumptions are satisfied in practice is another question.

\noindent\textbf{In implementation.}
Based on the ergodicity, we can implement the two error rates based on Eqs.(\ref{eq:pfa_range}-\ref{eq:pmiss_range}):
\begin{align}
     &\widehat{R}(\pi_\classtar, {\costfanon}, {\costmiss}) \\
    =&  {\costfanon}(1-\pi_\classtar) \widehat{P}_\text{fa}(\tau) + {\costmiss}\pi_\classtar \widehat{P}_\text{miss}(\tau)\\
    =& \frac{{\costfanon}(1-\pi_\classtar)}{T_\classnon} \int_{0}^{T_\classnon} \mathrm{1}(r(t) > \tau_\text{Bayes}) \, dt  \\
     &\quad +  \frac{{\costmiss}\pi_\classtar}{T_\classtar} \int_{0}^{T_\classtar} \mathrm{1}(r(t) < \tau_\text{Bayes}) \, dt
   \label{eq:bayesrisk:range_compute}
\end{align}
The computation recipe to compute the risk is:
\begin{enumerate}
    \item Given a large evaluation set, we concatenate the ground-truth fake and real segments into two long utterances with duration $T_\classnon$ and $T_\classtar$. We also concatenate the output of the models $r(t)$.
    \item Given a threshold $\tau$, we follow Eqs.(\ref{eq:pfa_range}-\ref{eq:pmiss_range}) (or the recipe in Range-EER paper) and compute $\widehat{P}_\text{fa}(\tau)$ and $\widehat{P}_\text{miss}(\tau)$.
    \item We plug them into Eq.(\ref{eq:bayesrisk:range_compute}) and compute $\widehat{R}(\pi_\classtar, {\costfanon}, {\costmiss})$.
\end{enumerate}

\subsection{Conclusion}
Range EER is linked to conventional binary classification evaluation but measures detection performance continuously across time. Under assumptions of stationarity and ergodicity in the score process, Range EER retains the theoretical foundation of conventional EER and represents the maximum decision risk under a uniform loss function with respect to the Bayes decision threshold.

The assumptions on the score process, the decision cost, and so on are introduced to link range EER with conventional EER. Future work has to re-think whether those assumptions are necessary or solid.

\end{document}